\documentstyle[12pt]{article}
\begin{document}
\vspace{0.5 cm}
\begin{center}
{\large \bf {\large Statistical Multifragmentation in Central Au+Au
Collisions
at 35 MeV/u}}
\end{center}
\vspace{0.5 cm}
\small
{M. D'Agostino$^{\$}$, A. S. Botvina$^{\ast}$, P. M. Milazzo$^{\&}$,
M. Bruno$^{\$}$, G. J. Kunde$^{\dag}$, D. R. Bowman$^{\ddag}$,
L. Celano$^{\#}$, N. Colonna$^{\#}$, J. D. Dinius$^{\dag}$,
A. Ferrero$^{\oplus}$, M. L. Fiandri$^{\$}$,
C. K. Gelbke$^{\dag}$, T. Glasmacher$^{\dag}$, F. Gramegna$^{\S}$,
D. O. Handzy$^{\dag}$, D. Horn$^{\ddag}$, W. C. Hsi$^{\dag}$,
M. Huang$^{\dag}$, I. Iori$^{\oplus}$, M. A. Lisa$^{\dag}$,
W. G. Lynch$^{\dag}$, L. Man\-du\-ci$^{\$}$, G. V. Margagliotti$^{\&}$,
P. F. Mastinu$^{\$}$, I. N. Mishustin$^{\wedge}$, C. P. Montoya$^{\dag}$,
A. Moroni$^{\oplus}$, G. F. Peaslee$^{\dag}$, F. Petruzzelli$^{\oplus}$,
L. Phair$^{\dag}$, R. Rui$^{\&}$, C. Schwarz$^{\dag}$, M. B. Tsang$^{\dag}$,
G. Vannini$^{\&}$, C. Williams$^{\dag}$}

\scriptsize
$^{\$}$ Dipartimento di Fisica and INFN, Bologna, Italy

$^{\ast}$ Hahn-Meitner-Institute, Berlin, Germany
 and Institute for Nuclear Research, Russian Academy of Science,
 117312 Moscow, Russia

$^{\&}$ Dipartimento di Fisica and INFN, Trieste, Italy

$^{\dag}$ NSCL, Michigan State University, USA

$^{\ddag}$ Chalk River Laboratories, Chalk River, Canada

$^{\#}$  INFN, Bari, Italy

$^{\oplus}$ Dipartimento di Fisica and INFN, Milano, Italy

$^{\S}$ INFN, Laboratori Nazionali di Legnaro, Italy

$^{\wedge}$ Niels Bohr Institute, DK-2100 Copenhagen, Denmark
 and Kurchatov Institute, Russian Scientific Center, 123182 Moscow, Russia

\vspace{0.5 cm}
\normalsize
\rm
\begin{abstract}
Multifragment disintegrations, measured for central $Au + Au$ collisions at
$E/A = 35 MeV$, are analyzed with the Statistical Multifragmentation Model.
Charge distributions, mean fragment energies, and two-fragment correlation
functions are well reproduced by the statistical breakup of a large, diluted
and thermalized system slightly above the multifragmentation threshold.
\end{abstract}

Hot nuclear systems produced in intermediate energy nucleus-nucleus
collisions
are known to decay by multiple fragment emission~\cite{1,2,3}. While this
decay mode may be related to a "liquid-gas" phase transition in finite
nuclear systems, both statistical and dynamical aspects of multifragmenting
finite nuclear systems must be understood before inferences about nuclear
phase
transitions can be made. Since the thermodynamic limit should be more
readily
reached for the heaviest possible nuclear systems, studies of central
collisions between heavy nuclei, such as $Au + Au$, are of particular
relevance.

Central collisions between heavy nuclei produce maximum fragment
multiplicities at incident energies of $E/A \approx 100 MeV$~\cite{4,5}.
Hot nuclear systems formed at this energy have, however, been shown to
undergo
a rapid collective expansion~\cite{6,7} which complicates a statistical
interpretation of the decay~\cite{8,9}. The dynamics of collective expansion
should be less important at lower incident energies~\cite{10,11} where
comparisons with equilibrium statistical model calculations~\cite{1,3}
should
be more appropriate. In this paper we perform such a comparison for
fragments
produced in central $Au + Au$ collisions at $E/A = 35 MeV$.

The experiment was performed at the National Superconducting Cyclotron
Laboratory of the Michigan State University. Experimental details have
already
been reported in Refs.~\cite{10,11}. Briefly, charged particles of
element number $Z \le 20$ were detected at $23^o \le \theta_{lab} \le 160^o$
by 171 phoswich detector elements of the MSU Miniball array~\cite{12};
fragments with charge up to $Z = 83$  were detected at
$3^o \le \theta_{lab} < 23^o$ by the {\it Multics} array~\cite{13}. The
geometric acceptance of the combined array was greater
than 87\% of $4 \pi$. The charge identification thresholds in
the Miniball were $E_{th}/A \approx 2, 3, 4 MeV$ for $Z = 3, 10, 18$,
respectively, and $E_{th}/A \approx 1.5 MeV$ in the Multics array,
independent of the fragment charge. For the present analysis, central
collisions are selected by requiring observed charged particle
multiplicities
$N_c > 24$ (representing about 10\% of the total reaction cross
section~\cite{11}).

For central events, the fragment emission was found~\cite{10,11}
compatible with a near-isotropic decay of a source consisting of more than
300
nucleons and with negligible contributions from the decay of projectile and
target-like residues. Here we address the question to which degree
fragment emission can be described by a statistical equilibrium calculation
performed with the Statistical Multifragmentation Model (SMM) of
Refs.~\cite{3,14}. This model was successfully applied at higher energies to
the interpretation of $Au$-induced projectile fragmentation
reactions~\cite{15}
for which collective expansion velocity components have been shown to be
negligible and also for central collisions~\cite{16} for which the model has
to be modified to take into account the large observed expansion velocities.

The model is based upon the assumption of statistical equilibrium at a
low-density freeze-out stage of the reaction at which the primary fragments
are formed according to their equilibrium partitions. The equilibrium
partitions are calculated according to the micro-canonical ensemble of all
break-up channels composed of nucleons and excited fragments of different
masses~\cite{16}. The model conserves total excitation energy, momentum,
mass and charge number. The statistical weight of decay channel j is given
by
$W_{j} \propto exp~S_{j}(E_s^{*},A_s,Z_s)$, where $S_{j}$ is the entropy of
the system in channel $j$ and $E_s^{*}, A_s$, and $Z_s$ are the excitation
energy, mass and charge number of the source. Different breakup
configurations
are initialized according to  their statistical weights. The fragments are
then propagated in their mutual Coulomb field and allowed to undergo
secondary
decays.
Light fragments with mass number $A_f\leq 4$ are considered as stable
particles ("nuclear gas") with only translational degrees of freedom;
fragments with $A_f > 4$ are treated as heated nuclear liquid  drops.
The secondary decay of large fragments ($A_f > 16$) is calculated from an
evaporation-fission model, and that of smaller fragments from a Fermi
break-up model~\cite{14}. The simulated events were then filtered with the
acceptance of the experimental apparatus~\cite{11,17}.
The same normalization to the total number of events was applied to
the experimental and calculated distributions, which thus may be compared
on an
absolute scale.

In its original version~\cite{14}, the SMM only incorporates thermal degrees
of freedom, i.e. the fragment energy distribution was determined from the
source temperature and then modified by final-state Coulomb interactions and
secondary decays. In the present work, we also allow for a collective radial
expansion of the system which could arise from a rapid thermal expansion,
possibly aided by an initial compression. Specifically, we assume that
modest
collective velocity components do not influence the fragment formation
probabilities for a given thermal energy; this assumption is reasonable for
$E_{flow}/A \leq 3 MeV$~\cite{3}. A self-similar collective expansion
was assumed, $v_{flow} \propto r$, where $r$ is the distance from the
source's center of mass. This collective velocity was added to the
thermal fragment velocity. The energy balance was taken into account.

In order to obtain an equilibrium freeze-out condition, we need to estimate
mass, charge, and energy carried away by particles of early emission. For
this
purpose, we assume that the preequilibrium emission consists
primarily of $n, p, d, t, ^{3}He$ and $\alpha$-particles distributed
uniformly in the available phase space in centre-of-mass system.
Following Ref.~\cite{3}, we search for parameters (mass $A_s$, charge $Z_s$,
excitation energy $E_s^{*}$, and freeze-out density $\rho_s$, assuming
$Z_s/A_s = 79/197$) of thermalized sources which can fit the
experimental data. A rough estimate of $A_s$, $Z_s$, and $E_s^{*}$
was obtained by comparing measured and predicted multiplicities of charged
particles, $N_c$, and intermediate mass fragments
$N_{IMF}$ ($3\le Z \le 30$). To explore sensitivities to the assumed
freeze-out density, we performed calculations for $\rho_s = \rho_0/3$ and
$\rho_0/6$ (where $\rho_0 \approx 0.15 fm^{-3}$ is the density of normal
nuclear matter).
Assuming $\rho_s = \rho_0/3$, the model can reproduce central collision data
for $Z_s \approx (0.7 - 0.9) Z_{tot}$ and
$E_s^{*} \approx (0.6 - 0.7) E_{tot}$
where $Z_{tot}$ and $E_{tot}$ denote the total available charge and
center-of-mass energy. For $\rho_s = \rho_0/6$, the source parameters are
10\%
smaller. The approximate temperature and entropy per nucleon of
the extracted source are $T \approx 6 MeV$ and $S/A \approx 1.5 - 1.6$.

Improved source parameters can be obtained by analyzing the inclusive
charge
distributions $N(Z) = Yield(Z)/N_{events}$ (Fig.~1~a,b) and the exclusive
charge distributions for the "first" six fragments ordered according to
$Z_i \ge Z_k$ if $i < k$ (Fig.~2).
We show SMM calculations for two sets of source parameters:
\begin{description}
\item $A_s=343$, $Z_s=138$, $E_s^{*}/A = 6.0 MeV$, $\rho_s = \rho_0/3$,
\item $A_s=315$, $Z_s=126$, $E_s^{*}/A = 4.8 MeV$, $\rho_s = \rho_0/6$.
\end{description}
In the second case an additional radial flow energy $E_{flow}/A = 0.8 MeV$
was introduced (the total energy of the source was thus $E_s/A = 5.6 MeV$).
This modest amount of radial flow is in agreement with the extrapolation of
data at higher energies to $E/A = 35 MeV$~\cite{18}. As shown below,
both sets of calculations well reproduce the measured
observables.
We have also considered a collective rotational motion with the same energy
per
nucleon as the flow energy and we found our main
conclusions to be essentially unchanged as compared to the case of
collective expansion.

In Fig.~1, the sensitivity to the excitation energy is illustrated by
additional calculations performed for $E_s^{*}/A \pm 1 MeV$
{\it per nucleon}
(dot-dashed and dotted curves).
A comparison of filtered and unfiltered calculations (solid and dashed
curves,
respectively) shows that distortions of the Z-distribution from to the
experimental apparatus are relatively small, except in the region $Z > 20$
where the Miniball detectors lose charge resolution. In the region of large
Z, the charge distribution falls off more steeply than expected for
exponential or power-law distributions. This steep fall-off, reproduced in
the
calculations, is an effect of charge (mass) conservation for a finite system.

One can readily understand why sources of different excitation energy and
density can produce similar fragment distributions. A system of lower
density
has smaller Coulomb barriers for fragment formation and thus requires a
smaller temperature to break up into fragments. During the expansion to a
lower density additional particles will be lost and some of the internal
energy may be converted into radial flow. Therefore, the low-density source
should have less mass and excitation energy.

Mean values, $\langle E/A \rangle$ (solid points), and standard deviations
$\Delta E/A$ (vertical bars) of the kinetic energies per nucleon of
fragments emitted at $\theta_{cm} = 90^o \pm 10^o$ are shown in Fig.~3 (left
panel) as a function of $Z$. The solid and dashed curves
show the results of SMM calculations at $\langle E/A \rangle \pm \Delta E/A$
for $\rho_s = \rho_0/3$ and $\rho_0/6$, respectively, filtered by the
acceptance of the experimental apparatus. Right panels of the
figure show the dependence of $\langle E/A \rangle$ on $\theta_{cm}$
for $Z = 6, 10$, and $14$. (The small rise in $\langle E/A \rangle$ at
forward angles and the small dip at backward angles are caused by the
acceptance of the experimental apparatus~\cite{11}.)
Overall, both SMM calculations reproduce the data rather well -- the smaller
Coulomb repulsion from the lower-density source is compensated by the added
collective expansion energy. Both calculations underpredict the kinetic
energies of lighter fragments ($Z < 5$) and of fragments emitted at forward
angles, possibly due to the presence of some nonequilibrium fragment
emission
in the data. The bulk of the data is, however, consistent with
near-equilibrium
emission from a single source~\cite{11}.

In order to test the predicted spatial separation of emitted fragments, we
have constructed two-fragment correlation functions~\cite{19},
$$1+ R(v_{red}) = C \frac{Y(v_{red})} {Y_{back}(v_{red})} $$
where
$$v_{red}  = \frac{ \mid \vec v_i - \vec v_j \mid} {\sqrt (Z_{i}+Z_j)}$$
is the "reduced" relative velocity of fragments i and j $(i \not= j)$
with charges $Z_i$ and $Z_j$; $Y(v_{red})$ and $Y_{back}(v_{red})$
are the coincidence and background yields for fragment pairs of reduced
velocity $v_{red}$; and $C = N_{back}/N_{coinc}$ where $N_{coinc}$ and
$N_{back}$ are the total number of coincidence and background pairs. The
background yield was constructed by means of the mixed event
techniques~\cite{20,21}.

Because of the large differences in dynamic range and resolution between the
Multics and Miniball arrays, we separately evaluated the two-fragment
correlation functions measured with the two devices. Figures 4~a) and
{}~b) show, respectively, two-fragment correlations constructed solely from
fragments detected in the Multics array
($3 \le Z \le 30$, $8^o\le \theta_{lab} < 23^o$)
and in the Miniball ($3 \le Z \le 10$, $23^o\le \theta_{lab} \le 40^o$).
The solid and dashed curves show the results of SMM calculations
performed for $\rho_s = \rho_0/3$ and $\rho_0/6$, respectively,
filtered for the acceptance of the experimental apparatus. Overall,
the measured correlation functions are approximately reproduced
by the calculations. Some discrepancies may be caused by the small
differences
found in the event charge partitions shown in Fig.~2 (see analysis on
Ref.~\cite{22}).
The increased overshoot just after the {\it Coulomb hole} for the case
$\rho_s = \rho_0/6$ is caused by the flow: more "organized" motion produces
additional correlations in single events.

Two-fragment correlation functions measured in this experiment are also
reproduced~\cite{11} by assuming sequential emission from the surface of a
spherical source of charge $Z_s = 138$, breakup density $\rho_s = \rho_0/4$,
radial expansion velocity $v_{flow} = 1.4 \ cm/ns$ and average interfragment
emission time $\tau \approx 85 fm/c$.
This ambiguity in interpretation is due to the fact that the angle
integrated
correlation functions are affected by both the source's size and its
lifetime,
resulting in the well known space-time ambiguity~\cite{19,23}.
However the energy and angle dependent analysis~\cite{23,24,25} indicates
that
fragment emission may be sufficiently fast to justify the approximation of
instantaneous emission. In the SMM we increase effectively the time of
fragment
production (and decrease the initial correlation) by including the secondary
de-excitation.

The source parameters extracted in our analysis are consistent with
expectation from dynamical simulations.
We have performed such simulations with the Boltzmann Nordheim Vlasov (BNV)
model~\cite{26} for central collisions ($b \le 1\ fm$) using the mean-field
approximation with a soft equation of state. Approximately $100 fm/c$ after
the
initial contact between projectile and target, the calculations predict the
formation of a single source of mass $A \approx 324$, charge
$Z \approx 136$,
density $\rho = \rho_0/2$, excitation energy per nucleon
$E^*/A \approx\ 6 MeV$, and rotational energy $E_{rot} \approx 10 MeV$.
Unfortunately, BNV-type models do not treat emission of single fragments.

In conclusion, fragment emission observed in central $Au+Au$ collisions
at $E/A
= 35 MeV$ is largely consistent with the statistical break-up of a single
source of excitation energy per nucleon $E_s^{*}/A \approx 5 \div 6 MeV$
and density $\rho_s = \rho_0/3 \div \rho_0/6$.
Collective expansion energy is not excluded but it is smaller than
$\sim 1 MeV/u$.
These conditions are comparable to those observed for peripheral projectile
fragmentation reactions at higher energies ($E/A > 400 MeV$).
However, these reactions can only produce smaller sources of mass and charge
numbers $A_s < A_{projectile}$ and $Z_s < Z_{projectile}$.

\vspace{1.0 truecm}
The authors would like to thank A. Bonasera and M. Di Toro for many
interesting and stimulating discussions and for making their codes available.
The technical assistance of R. Bassini, C. Boiano, S. Brambilla,
G. Busacchi,
A. Cortesi, M. Malatesta and R.Scardaoni during the measurements
is gratefully acknowledged.
This work has been supported in part by funds of the Italian Ministry
of University and Scientific Research and by the U.S. National Science
Foundation under Grant number PHY-92-14992. A.S. Botvina thanks the
Hahn-Meitner Institute for hospitality and support, I.N.Mishustin thanks
the Niels Bohr Institute
for hospitality and Carlsberg Foundation (Denmark) for financial support.

\newpage
\normalsize
\rm
{\bf {Figure captions}}\\
\small
\baselineskip=0.3 pt
{\bf Fig.1:} {Charge distribution $N(Z)$. Points show
experimental data and lines show results of SMM predictions for sources
with parameters $A_s = 343$, $Z_s = 138$, $E_s^{*}/A = 6.0 MeV$,
$\rho_s = \rho_0/3$ (part ~a)) and $A_s = 315$, $Z_s = 126$,
$E_s^{*}/A = 4.8 MeV$, $E_{flow}/A = 0.8 MeV$,
$\rho_s = \rho_0/6$ (part ~b)).
Dashed curves are the unfiltered calculations and solid curves are the
filtered ones. The dot-dashed and dotted curves represent filtered
calculations for thermal excitations $E_{s}^{*}/A + 1MeV/u$ and
$E_{s}^{*}/A - 1MeV/u$, respectively.}\\
{\bf Fig.2:} {Charge distributions of the six heaviest fragments in
each event, ordered such as $Z_i \ge Z_k$ if $i < k$.
Experimental data are shown by points, the solid and dashed curves show the
results of SMM calculations for $\rho_s = \rho_0/3$, and
$\rho_s = \rho_0/6$,
respectively (other source parameters as in Fig. 1).}\\
{\bf Fig.3:} { Mean centre-of-mass kinetic energy per nucleon,
$\langle E/A \rangle$, as a
function of the charge Z, for fragments emitted at
$\theta_{cm} = 90^o \pm 10^o$ (left panel) and (for $Z = 6, 10, 14$) as a
function of $\theta_{cm}$ (right panels).
Points give the experimental values of $\langle E/A \rangle$ and vertical
bars
give the standard deviations $\Delta E/A$ of the distributions.
The solid and dashed lines are SMM predictions of $\langle E/A \rangle$
(in the left panel $ \pm  \Delta E/A$) for $\rho_s = \rho_0/3$, and
$\rho_s = \rho_0/6$, respectively (other source parameters as in Fig. 1).
The energy range is the same in the left and in each right panel.}\\
{\bf Fig.4:} {Two-fragment correlation functions $1 + R(v_{red})$
for $3 \le Z \le 30 $ and $8^o \le \theta_{lab} < 23^o$ (part~a)) and
for $3 \le Z \le 10 $ and $23^o \le \theta_{lab} \le 40^o$ (part ~b)).
Full points show experimental data. The solid and dashed lines are SMM
predictions for $\rho_s = \rho_0/3$, and $\rho_s = \rho_0/6$,
(other source parameters as in Fig. 1).}
\end{document}